\documentclass[usenatbib]{mn2e}
\usepackage{lscape}
\usepackage{graphicx,graphics}
\usepackage{amssymb,amsmath}
\usepackage{natbib}
\usepackage[top= 2cm,bottom=2cm,left=2cm,right=2cm]{geometry}

\begin{document}

\title{Empirical $\lowercase{ugri}-UBVR_c$ Transformations for Galaxies}

\author[D. Cook et al.]
{David O. Cook,$^1$
Daniel A. Dale$^1$
Benjamin D. Johnson$^2$
Liese Van Zee$^3$
\newauthor
Janice C. Lee$^4$
Robert C. Kennicutt$^{5,6}$
Daniela Calzetti$^7$
Shawn M. Staudaher$^1$
\newauthor
Charles W. Engelbracht$^{8}$\\
$^1$Department of Physics \& Astronomy, University of Wyoming, Laramie, WY 82071, USA; dcook12$@$uwyo.edu\\
$^2$UPMC-CNRS, UMR7095, Institut d'Astrophysique de Paris, F-75014, Paris, France\\
$^3$Department of Astronomy, Indiana University, Bloomington, IN 47405, USA\\
$^4$Space Telescope Science Institute, 3700 San Martin Drive, Baltimore, MD 21218, USA\\
$^5$Institute of Astronomy, University of Cambridge, Cambridge CB3 0HA, UK\\
$^6$Steward Observatory, University of Arizona, Tucson, AZ 85721, USA\\
$^7$Department of Astronomy, University of Massachusetts, Amherst, MA 01003, USA\\
$^8$Raytheon Company, 1151 East Hermans Road, Tucson, AZ 85756, USA; Deceased}

\maketitle

\begin{abstract}
We present empirical color transformations between Sloan Digital Sky Survey $ugri$ and Johnson-Cousins $UBVR_c$ photometry for nearby galaxies ($D < 11~$Mpc). We use the Local Volume Legacy (LVL) galaxy sample where there are 90 galaxies with overlapping observational coverage for these two filter sets. The LVL galaxy sample consists of normal, non-starbursting galaxies. We also examine how well the LVL galaxy colors are described by previous transformations derived from standard calibration stars and model-based galaxy templates. We find significant galaxy color scatter around most of the previous transformation relationships. In addition, the previous transformations show systematic offsets between transformed and observed galaxy colors which are visible in observed color-color trends. The LVL-based $galaxy$ transformations show no systematic color offsets and reproduce the observed color-color galaxy trends. 
\end{abstract}


\section{Introduction} 
The Sloan Digital Sky Survey (SDSS) is a large-area optical survey which covers more than one quarter of the sky \citep{york00}. The five filters ($ugriz$) developed for this survey are based on a modified Thuan-Gunn \citep{ThuanGunn76} system which provide wide optical wavelength coverage and little overlap between each filter bandpass. Due to the wide acceptance of this filter system, an increasing number of ground-based surveys utilize SDSS filters. Thus there is a large amount of optical data for stars, galaxies, and quasars based on $ugriz$ photometry in addition to traditional $UBVR_cI_c$ photometry \citep{Johnson53,Cousins76}.  

All previous SDSS data releases have empirically characterized SDSS-Johnson-Cousins transformations for quasars \citep[e.g.,][]{jester05} and stars \citep[Pop I and II, metal-poor, dwarfs, and main sequence;][]{smith02,west05,karaali05,jester05,rodgers06,jordi06}. However, no study to date has explicitly investigated empirical transformations for a population of galaxies. It is reasonable to assume that some stellar transformations would describe galaxy transformations with relative accuracy since the majority of a galaxy's optical light emanates from stars. However, previous studies have found different transformations for stars with different metallicities \citep{karaali05,jordi06} and different colors \citep{smith02,jester05,west05,jordi06}. In addition, nebular emission and absorption are likely to significantly affect galaxy optical fluxes. It is not obvious if the transformations of stars with similar properties (i.e., metallicity and optical color) would accurately describe a transformation of an amalgamation of stars with a wide range of properties (i.e., a galaxy) and account for nebular emission and absorption.

Although no previous study has investigated $empirical$ galaxy transformations, the study of \citet{blanton07} has characterized transformations derived from model-based galaxy templates. These templates are a linear combination between observed galaxy spectral energy distributions (SEDs) from large galaxy surveys (GALEX, SDSS, 2MASS, GOODS, DEEP2) and the stellar models of \citet{bruzual03}. Although these transformations were derived for galaxies, they are dependent upon stellar models which introduce uncertainty based upon model assumptions (e.g., age, metallicity, IMF, stellar mass-to-light ratios) in addition to assumptions of a galaxy's star formation history. It is not clear if transformations derived from model-based galaxy templates will accurately describe transformations for observed galaxy colors.

This study utilizes the Local Volume Legacy \citep[LVL][]{dale09} survey galaxy sample to derive empirical SDSS-Johnson-Cousin transformations explicitly for galaxies. The LVL sample contains 90 galaxies with overlapping $UBVR_c$ and $ugriz$ observational coverage. This data set is ideal to characterize these transformations since the LVL sample consists of galaxies with a wide range of properties (e.g., morphology, metallicity, optical luminosity and color, star formation rate, etc.). Furthermore, the sample is dominated by low-mass galaxies with low internal extinction \citep[][Cook et al. 2014c]{dale09}. These data in combination with previous stellar transformation relationships will facilitate an examination of how well stellar transformations describe galaxy transformations. In addition, we compare the LVL galaxy transformations to the \citet{blanton07} transformations derived from model-based galaxy templates.

\section{Sample \label{sec:sample}}
The LVL sample consists of 258 of our nearest galaxy neighbors reflecting a statistically complete, representative sample of the local universe. The sample selection and description are detailed in \citet{dale09}, but we provide an overview here. 

The LVL sample was built upon the samples of two previous nearby galaxy surveys: ACS Nearby Galaxy Survey Treasury \citep[ANGST;][]{dalcanton09} and 11~Mpc H$\alpha$ and Ultraviolet Galaxy Survey \citep[11 HUGS;][]{kennicutt08,lee11}. The final LVL sample consists of galaxies that appear outside the Galactic plane ($|b| > 20^{\circ}$), have a distance less than 11~Mpc ($D \leq$ 11~Mpc), span an absolute $B-$band magnitude of $-9.6 < M_B < -20.7$ mag, and span an RC3 catalog galaxy type range of $-5 < T < 10$. Consequently, the LVL sample contains low-mass dwarf and irregular galaxies, large spirals, and a few elliptical galaxies. 

\begin{figure*}
  \begin{center}
  \includegraphics[scale=0.8]{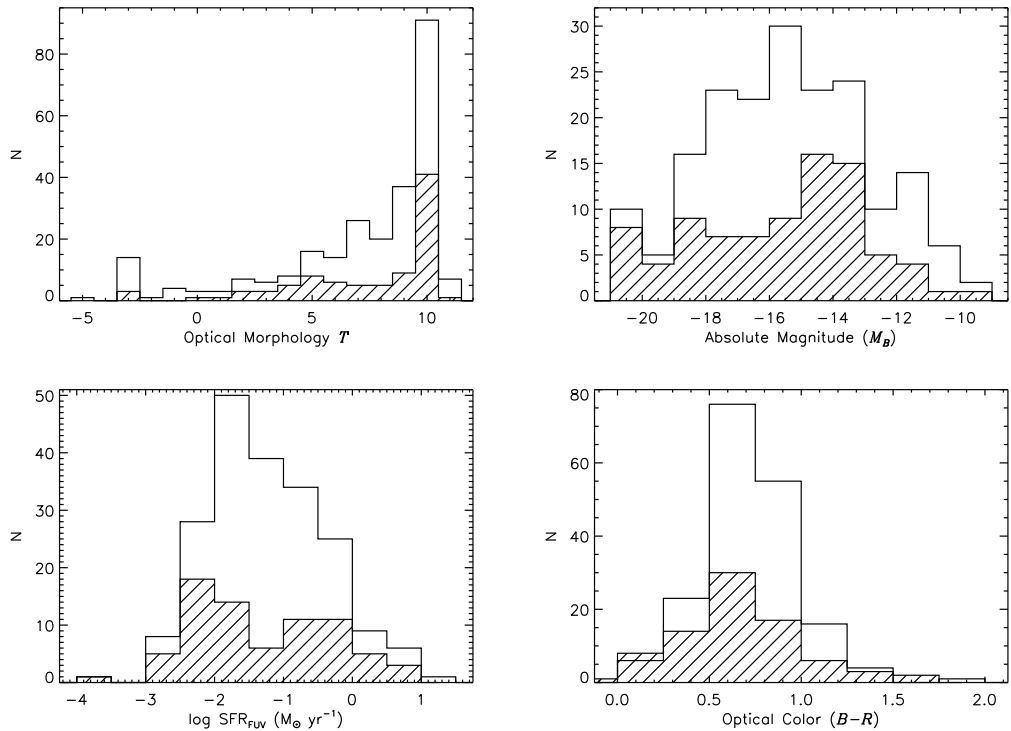}
  \caption{Distributions of RC3 morphology type taken from \citet{kennicutt08} (top left), absolute $B-$band magnitude (top right), star formation rate (bottom left), and $(B-V)$ optical color (bottom right). The unshaded histograms show the entire sample of 258 galaxies, whereas the shaded portions indicate a sub-sample of galaxies with overlapping $UBVR_c$ and $ugri$ observations.}
   \label{fig:sample}
   \end{center}
\end{figure*}

There is significant $UBVR_c$ and $ugriz$ observational overlap. A total of 49, 85, 39, and 88 galaxies with $UBVR_c$ observations, respectively, also have $ugriz$ imaging available. Figure~\ref{fig:sample} presents histograms of the full LVL sample (unshaded) and a sub-sample with overlapping $ugriz$ observational coverage (shaded). Both samples show relatively similar distributions of morphology, absolute $B-$band magnitude, star formation rate (SFR), and $B-R$ optical color. The absolute $B-$band magnitudes and $B-R$ optical colors are taken from Cook et al. (2014a; submitted). The SFRs are derived from the GALEX far-ultraviolet (FUV) magnitudes of \citet{lee11} and are transformed into SFRs via the prescription of \citet{murphy11}.

\section{Data \label{sec:data}}
We utilize the $UBVR_c$ and $ugri$ fluxes from the LVL global optical photometry study of Cook et al. (2014a; submitted). We do not use the $z-$band fluxes for our transformations since there are no corresponding $I_c-$band fluxes. These data are fully described in Cook et al. (2014a), but we provide a brief overview here. 

The $UBVR_c$ ground-based data were taken from 1--2 meter class telescopes. The images were reduced via standard IRAF tasks and the resulting photometry was calibrated to the Johnson-Cousins magnitude system via Landolt standard star observations taken several times a night. Since the fluxes of Cook et al. (2014a) are published in the AB magnitude system, we have recovered the Vega magnitudes via the prescription of \citet{blanton07}.

The SDSS data were downloaded from SDSS DR7 \citep[][]{sdss7}. For each galaxy, mosaic images were constructed using the utility SWARP \citep{swarp}, sky-subtracted, and photometrically calibrated to the AB magnitude system; note that the SDSS magnitudes are corrected for AB magnitude offsets as prescribed by the DR7 data release website.\footnote{http://www.sdss.org/dr7/algorithms/fluxcal.html} 

To ensure accurate photometric comparisons across different optical bandpasses, Cook et al. (2014a; submitted) removed contaminating foreground stars and background galaxies from the optical images. Contaminant regions were overlaid onto each optical image, visually inspected, and the contaminant region sizes were adjusted to account for the relative apparent brightness of each source. Each contaminant was removed through an interpolation of the surrounding local sky using the IRAF task IMEDIT.

Global photometry is performed on both $UBVR_c$ and $ugri$ images within identical apertures defined by \citet{dale09}. These apertures were chosen to encompass the majority of the emission seen at GALEX UV (1500\AA--2300\AA) and Spitzer IR (3.6$\mu$m--160$\mu$m) wavelengths, but we have visually checked to ensure that they are adequately large to encompass all of the optical emission of each galaxy. These apertures have a median semi-major axis ratio of 1.5 compared to R25 apertures \citep{rc3}. 

\section{Results \label{sec:results}}
In this section we derive empirical color transformations for galaxies using a sub-sample of LVL with overlapping $UBVR_c$ and $ugri$ imaging. We derive both $ugri$-to-$UBVR_c$ and $UBVR_c$-to-$ugri$ transformations and quantify the accuracy of all possible color-color combinations via the RMS scatter and the resulting transformed minus observed magnitude for each color transformation. We compare these results to previous transformations and examine how well all transformations reproduce observed color-color trends. The LVL galaxy transformations show better agreement with observed magnitudes and colors compared to previous transformations.

\subsection{SDSS to Johnson-Cousins \label{sec:SDSStoJohn}}
In this section we describe the methods used to derive the $ugri$-to-$UBVR_c$ galaxy transformations. We quantify the accuracy of each transformation and compare them to previous transformations. In addition, we perform a second accuracy check by comparing transformed colors to observed color-color trends. 

\subsubsection{LVL Transformations for Galaxies \label{sec:LVLtrans}}
To determine color transformations between the SDSS and Johnson-Cousins magnitude systems we examine all possible combinations of the difference in Johnson-Cousins and SDSS magnitudes versus SDSS colors (e.g., $U-r$ versus $g-r$). A least $\chi^2$ fit is performed on each combination and the RMS scatter about the best-fit line is calculated. In addition, we calculate the median absolute difference between the transformed and observed magnitude ($\Delta M_{\rm{med}}$) for each color transformation. 

The combination of RMS scatter and $\Delta M_{\rm{med}}$ values together quantify the accuracy of each color transformation. A low RMS scatter implies a well-behaved transformation (i.e., most galaxies show agreement with the best fit transformation). The $\Delta M_{\rm{med}}$ values quantify how accurately the transformation (i.e. the best-fit line to the data) reproduces the typical observed magnitudes for the sample. 

\begin{table}
\begin{center}
  {SDSS Transformations}
\end{center}
\begin{tabular}{cccc}
\hline
\hline
Color & & Color Transformation & RMS  \\
      & & 			 & (mag) \\
\hline
$U-i$ & = & $(~~1.93 \pm 0.06)(g-i) + (-0.48 \pm 0.04)$ & $0.09$\\
$B-i$ & = & $(~~1.27 \pm 0.03)(g-i) + (~~0.16 \pm 0.01)$ & $0.04$\\
$V-u$ & = & $(-0.82 \pm 0.03)(u-r) + (-0.02 \pm 0.04)$ & $0.03$\\
$R-u$ & = & $(-1.06 \pm 0.02)(u-r) + (-0.11 \pm 0.03)$ & $0.08$\\
\hline
\end{tabular} \\
\caption{The $ugri$-to-$UBVR_c$ transformation equations and coefficients for galaxies derived from a sub-sample of LVL where there exists an overlap between $UBVR_c$ and $ugri$ imaging. The last column lists the RMS scatter of the LVL galaxy colors around the least $\chi^2$ fit for each transformation. }
\vspace{0.1 cm}
\label{tab:sdss2vega}
\end{table}

\begin{figure}
  \begin{center}
  \includegraphics[scale=0.5]{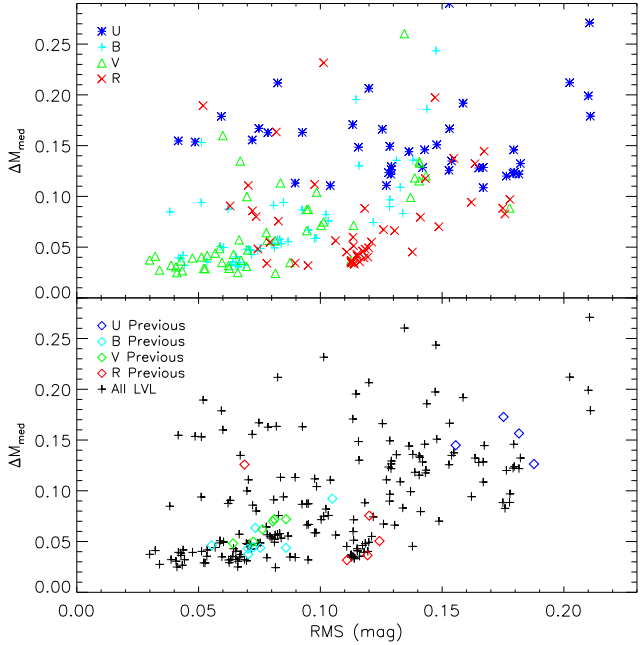}
  \caption{The median absolute difference between the transformed and observed magnitudes ($\Delta M_{\rm{med}}$) versus the RMS scatter of each color transformation. Top panel: all possible color-color transformation combinations of the LVL sub-sample with overlapping $UBVR_c$ and $ugri$ imaging. Bottom panel: transformations of previous studies derived from standard calibration stars overplotted onto the data of the top panel which are now presented as ``plus" signs.}
  \label{fig:sdss2vegarms}
   \end{center}
\end{figure}

The top panel of Figure~\ref{fig:sdss2vegarms} shows $\Delta M_{\rm{med}}$ versus the RMS scatter of each LVL color transformation. Above an RMS value of $\sim0.12$ magnitudes, there exists a lower envelope in $\Delta M_{\rm{med}}$ where the transformed magnitudes are increasingly inconsistent with the observed magnitude (i.e., higher $\Delta M_{\rm{med}}$). This trend suggests that transformations with low RMS scatter should result in good agreement between transformed and observed magnitudes. However, there are a number of color transformations with low RMS scatter ($<0.12$ mag) but relatively high $\Delta M_{\rm{med}}$ values ($>$0.05 mag). Thus a combination of low RMS scatter and low $\Delta M_{\rm{med}}$ would provide the best available transformation for each filter. If we treat the RMS scatter and $\Delta M_{\rm{med}}$ values as a measure of each transformation's uncertainty, we can add these quantities in quadrature, identify the lowest value as the transformation with the lowest uncertainty, and thus identify the best available transformation.

There are several $B-$ and $V-$band transformations where both the RMS scatter and $\Delta M_{\rm{med}}$ values fall below 0.05 magnitudes. Also, the $R_c-$band transformations show low $\Delta M_{\rm{med}}$ values ($\Delta M_{\rm{med}} < 0.05$) but slightly higher RMS scatter ($0.05 < RMS < 0.12$). For these three bandpasses ($BVR_c$), we choose the transformation with the lowest RMS and $\Delta M_{\rm{med}}$ values added in quadrature. 

\begin{figure}
  \begin{center}
  \includegraphics[scale=0.35]{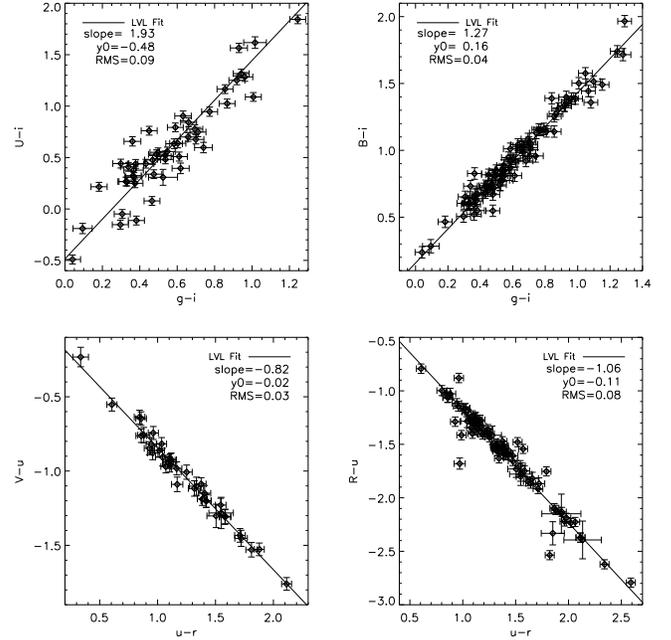}
  \caption{The color-color relationships used to define the LVL galaxy transformations. The solid black line is the least $\chi^2$ fit to the galaxy colors and represent the LVL galaxy transformations. The fit parameters are listed in the caption as well as in Table~\ref{tab:sdss2vega}.}
  \label{fig:lvltrans}
   \end{center}
\end{figure}

There are several $U-$band transformations which show low RMS scatter ($<0.12$ mag) but all show relatively high $\Delta M_{\rm{med}}$ values ($\geq$0.1 mag) compared to other bandpasses. The transformation with the lowest RMS and $\Delta M_{\rm{med}}$ value added in quadrature falls on Figure~\ref{fig:sdss2vegarms} at an RMS of $\sim0.09$ and $\Delta M_{\rm{med}}$ value of $\sim0.1$ magnitudes, respectively. The equations and best fit parameters of the $ugri$-to-$UBVR_c$ LVL galaxy transformations are shown in Table~\ref{tab:sdss2vega}. The individual LVL color transformation relationships are graphically presented in Figure~\ref{fig:lvltrans}.

\subsubsection{Previous Transformations \label{sec:prevtrans1}}
We also test how well the LVL galaxy colors are described by previous SDSS transformations. The studies of \citet{jester05}, \citet{jordi06}, and \citet{blanton07} derived $ugriz$-to-$UBVR_cI_c$ transformations while R. Lupton's (2005, hereafter L05)\footnote{https://www.sdss3.org/dr10/algorithms/sdssUBVRITransform.php} transformations derived $ugriz$-to-$BVR_cI_c$ transformations; there are no $U-$band transformations for L05. The studies of L05, \citet{jester05}, and \citet{jordi06} utilized either Landolt standard stars \citep{landolt92} or an extension of Landolt's standard stars \citep{stetson00} to derive stellar transformations, while \citet{blanton07} utilized model-based galaxy templates to derive galaxy transformations.

\begin{figure*}
  \begin{center}
  \includegraphics[scale=0.7]{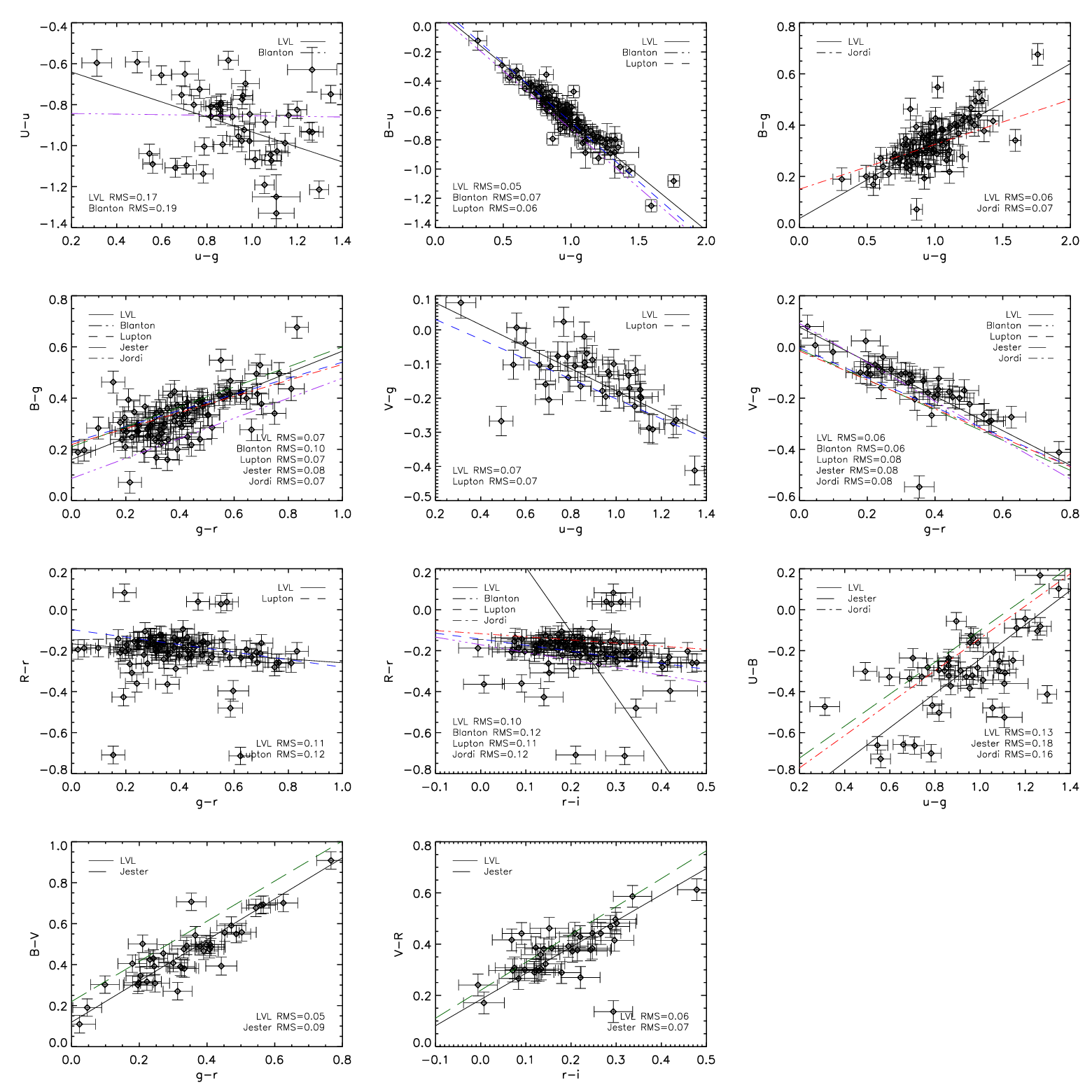}
  \caption{The color-color relationships used to derive the previous transformations of \citet{jester05}, L05, \citet{jordi06}, and \citet{blanton07}. The open diamonds represent the observed LVL galaxy colors and the black solid line represents the best-fit line to the data. The dashed, long-dashed, dashed-dot, and dashed-dot-dot lines represent the best-fit lines published by the studies of \citet{jester05}, L05, \citet{jordi06}, and \citet{blanton07}, respectively. Most panels show significant galaxy scatter or an offset between the LVL and stellar transformation best-fit line.}
  \label{fig:prevtrans}
   \end{center}
\end{figure*}

Figure~\ref{fig:prevtrans} graphically presents the color transformation relationships of these previous studies with the LVL galaxy colors overplotted. Any previous transformation which utilizes $I-$band photometry has been omitted since the study of Cook et al. (2014a; submitted) did not publish such measurements. The LVL galaxy colors show a significant amount of scatter around most of these relationships and/or there exists an offset between the LVL derived best-fit and those derived from standard stars and model-based galaxy templates. 

We calculate the RMS scatter of the LVL galaxy colors around the previously published best fit lines and the $\Delta M_{\rm{med}}$ values to quantify the accuracy of the previous transformations on galaxy colors. The results are graphically show in the bottom panel of Figure~\ref{fig:sdss2vegarms}. The previous transformations show a similar trend across $UBVR_c$ filters to those of LVL where the transformations of both the $B-$ and $V-$band show low RMS scatter and $\Delta M_{\rm{med}}$ values, the $R_c-$band show higher RMS scatter but low $\Delta M_{\rm{med}}$ values, and the $U-$band show relatively high RMS scatter and $\Delta M_{\rm{med}}$ values. 

Although some of the $B-$, $V-$, and $R_c-$band transformations of previous studies provide reasonably low $\Delta M_{\rm{med}}$ values, most of the corresponding color-color relationships show poor fits to observed galaxy colors. In addition, Figure~\ref{fig:sdss2vegarms} shows that for every transformation derived in previous studies there exists an LVL transformation with a lower RMS scatter and $\Delta M_{\rm{med}}$ value in every bandpass. Furthermore, the $U-$band transformations provide the highest contrast between the accuracies of the LVL and previous transformations. There are many LVL $U-$band transformations which are significantly lower in both RMS scatter and $\Delta M_{\rm{med}}$ values compared to the best $U-$band transformation derived from previous studies. 

\subsubsection{Observed Color-Color Trends}
The RMS scatter and $\Delta M_{\rm{med}}$ values provide one way to quantify the accuracy of each transformation. Another means to test the accuracy of color transformations is to propagate the transformed magnitudes into colors and examine how well these transformed colors reproduce observed color-color trends.

In general, galaxies show correlations between different optical colors  with varying amounts of scatter \citep[e.g., ][Cook et al. 2014a; submitted]{blanton03}. This scatter is reduced when the flux within each bandpass is corrected for internal extinction due to dust (Cook et al. 2014c; submitted). To test how well the LVL and previous transformations reproduce observed color-color relationships we correct both the transformed and observed LVL galaxy fluxes for extinction due to dust. 

The extinction corrections in each optical bandpass are carried out by Cook et al. (2014c; submitted) and are fully described there. Note that correcting for internal dust extinction does not change the difference between transformed and observed colors, but visually highlights any differences in color-color trends which are consequently tighter due to dust corrections. Quantities in this study which have been corrected for internal extinction due to dust are denoted with a subscript ``0" and represent an intrinsic measurement.

\begin{figure}
  \begin{center}
  \includegraphics[scale=0.57]{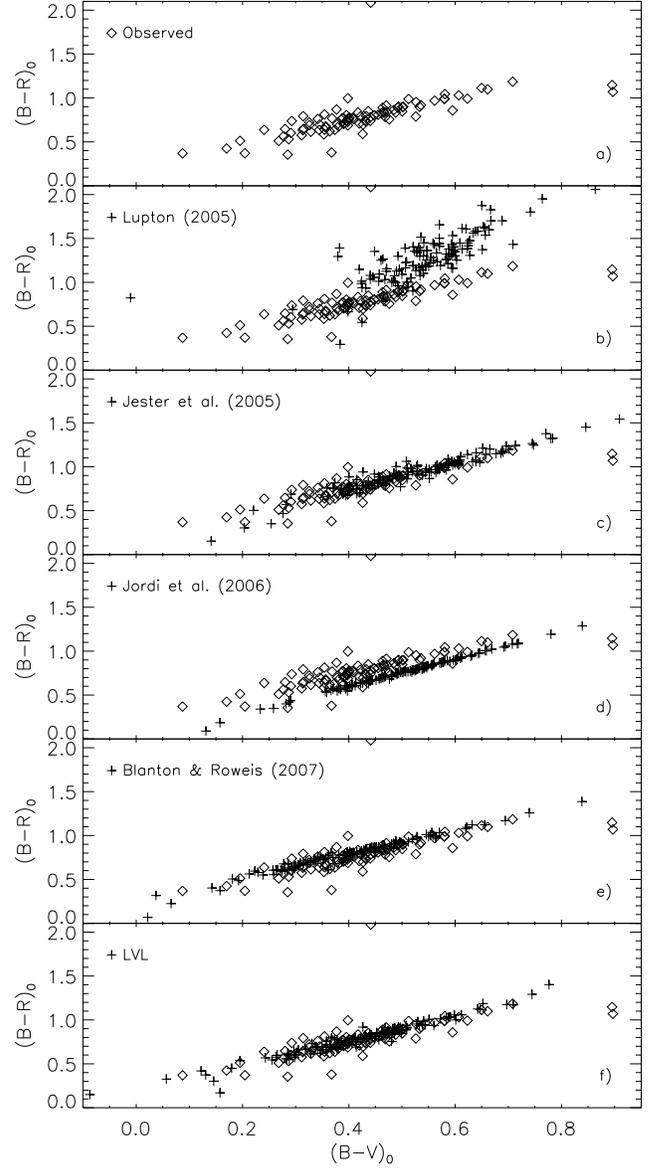}
  \caption{The color-color relationship between $(B-R)_0$ and $(B-V)_0$ where the flux of each bandpass in both colors have been corrected for internal extinction due to dust and Milky Way foreground extinction. Panel (a) is the observed trend with no transformed colors overplotted, panel (b) has the transformed colors via the prescription of L05 overplotted, panel (c) has the transformed colors via the prescription of \citet{jester05} overplotted, panel (d) has the transformed colors via the prescription of \citet{jordi06} overplotted, panel (e) has the transformed colors via the prescription of \citet{blanton07} overplotted, and panel (f) has the LVL transformed colors overplotted. Each of the stellar transformed colors show color offsets in both axes. These offsets are quantified in Table~\ref{tab:sdss2vegaDelColor}. The LVL transformed colors show good agreement with the observed color trend.}
  \label{fig:BVvBR}
   \end{center}
\end{figure}

\begin{figure}
  \begin{center}
  \includegraphics[scale=0.49]{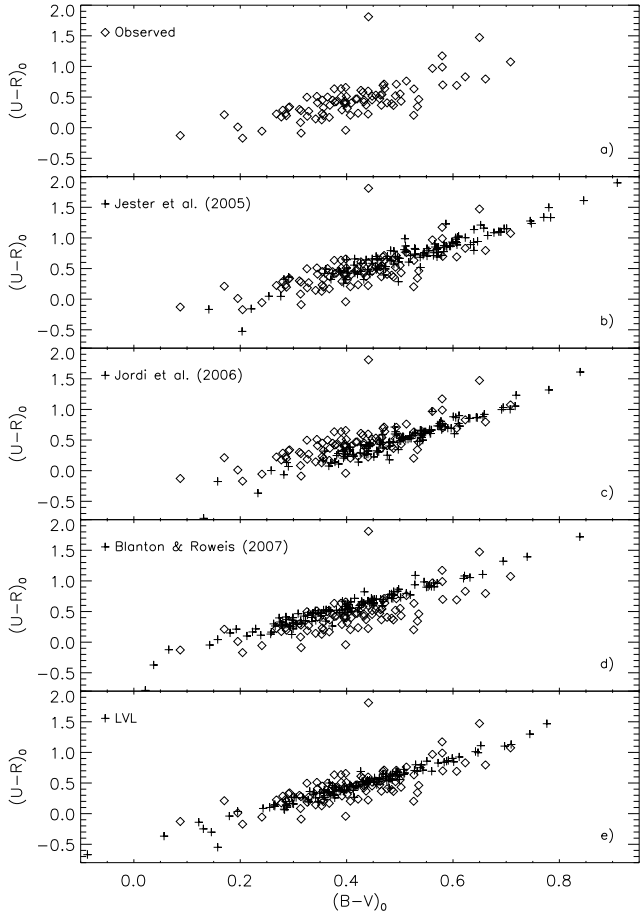}
  \caption{The color-color relationship between $(U-R)_0$ and $(B-V)_0$. Panel (a) is the observed trend with no transformed colors overplotted, panel (b) has the transformed colors via the prescription of \citet{jester05} overplotted, panel (c) has the transformed colors via the prescription of \citet{jordi06} overplotted, panel (d) has the transformed colors via the prescription of \citet{blanton07} overplotted, and panel (e) has the LVL transformed colors overplotted. Each of the stellar transformed colors show color offsets in both axes. These offsets are quantified in Table~\ref{tab:sdss2vegaDelColor}. The LVL transformed colors show good agreement with the observed color trend.}
  \label{fig:BVvUR}
   \end{center}
\end{figure}

Figures~\ref{fig:BVvBR} and \ref{fig:BVvUR} show two color-color relationships where the observed $UBVR_c$ colors show relatively tight correlations with each other. The panels of Figures~\ref{fig:BVvBR} and \ref{fig:BVvUR} present the observed-only color-color trend for a visual reference and subsequent comparison panels between observed and transformed colors. These comparison panels show colors transformed via the prescriptions of L05, \citet{jester05}, \citet{jordi06}, \citet{blanton07}, and the LVL galaxy prescriptions presented in Table~\ref{tab:sdss2vega}. When multiple previous transformations exist for a given bandpass, we choose the one which yielded the lowest $\Delta M_{\rm{med}}$ value in the bottom panel of Figure~\ref{fig:sdss2vegarms}.

Figure~\ref{fig:BVvBR} shows the $(B-V)_0$ versus $(B-R)_0$ relationship where panel (a) is the observed trend with no transformed colors overplotted, panel (b) has the transformed colors via the prescription of L05 overplotted, panel (c) has the transformed colors via the prescription of \citet{jester05} overplotted, panel (d) has the transformed colors via the prescription of \citet{jordi06} overplotted, panel (e) has the transformed colors via the prescription of \citet{blanton07} overplotted, and panel (f) has the LVL transformed colors overplotted.  

\begin{table}
\begin{center}
  {Median Color Differences}
\end{center}
\begin{tabular}{clc}
\hline
\hline
Color & Study & $\Delta color$ \\
      &       & (mag) \\
\hline
$(U-B)_0$.................... & Jester et al. (2005): & $~~0.15$\\
                              & Jordi~ et al. (2006): & $~~0.11$\\
                              & Blanton \& Roweis (2007): & $~~0.13$\\
                              & LVL:~~~~~~~~~~~~~~~ & $~~0.04$\\
\hline \vspace{-0.2 cm}\\
$(U-V)_0$.................... & Jester et al. (2005): & $~~0.29$\\
                              & Jordi~ et al. (2006): & $~~0.23$\\
                              & Blanton \& Roweis (2007): & $~~0.14$\\
                              & LVL:~~~~~~~~~~~~~~~ & $~~0.06$\\
\hline \vspace{-0.2 cm}\\
$(U-R)_0$.................... & Jester et al. (2005): & $~~0.24$\\
                              & Jordi~ et al. (2006): & $~~0.02$\\
                              & Blanton \& Roweis (2007): & $~~0.06$\\
                              & LVL:~~~~~~~~~~~~~~~ & $-0.01$\\
\hline \vspace{-0.2 cm}\\
$(B-V)_0$.................... & Jester et al. (2005): & $~~0.10$\\
                              & Lupton (2005):~~~~~~& $~~0.13$\\
                              & Jordi~ et al. (2006): & $~~0.08$\\
                              & Blanton \& Roweis (2007): & $-0.03$\\
                              & LVL:~~~~~~~~~~~~~~~ & $~~0.01$\\
\hline \vspace{-0.2 cm}\\
$(B-R)_0$.................... & Jester et al. (2005): & $~~0.16$\\
                              & Lupton (2005):~~~~~~& $~~0.50$\\
                              & Jordi~ et al. (2006): & $-0.01$\\
                              & Blanton \& Roweis (2007): & $~~0.01$\\
                              & LVL:~~~~~~~~~~~~~~~ & $~~0.01$\\
\hline \vspace{-0.2 cm}\\
$(V-R)_0$.................... & Jester et al. (2005): & $~~0.05$\\
                              & Lupton (2005):~~~~~~& $~~0.31$\\
                              & Jordi~ et al. (2006): & $-0.10$\\
                              & Blanton \& Roweis (2007): & $~~0.03$\\
                              & LVL:~~~~~~~~~~~~~~~ & $-0.01$\\
\hline
\end{tabular} \\
\caption{The median color differences between the $ugri$-to-$UBVR_c$ transformed and observed colors for the LVL galaxy transformations and those derived in previous transformations. The flux of each bandpass for all colors have been corrected for internal extinction due to dust.}
\label{tab:sdss2vegaDelColor}
\end{table}

Visual inspection of Figure~\ref{fig:BVvBR} shows significant disagreement between the observed color-color trend and the transformed colors of L05 in both slope and color offset. The transformed colors of \citet{jester05}, \citet{jordi06}, and \citet{blanton07} show similar slopes to that of the observed trend, but the \citet{jordi06} transformations show a noticeable offset. 

Although the transformed colors of \citet{jester05} show a similar slope to that of the observed colors there is an offset which is less obvious at first glance. Table~\ref{tab:sdss2vegaDelColor} shows the median color difference between the transformed and observed colors for the LVL and previous transformations. The median $\Delta(B-R)_0$ and $\Delta(B-V)_0$ color differences for \citet{jester05} show that both colors are redder by 0.16 and 0.1 magnitudes, respectively. The ratio of $\Delta(B-R)_0$-to-$\Delta(B-V)_0$ represents the slope at which the \citet{jester05} colors are shifted and is similar to the slope of the observed color-color trend (ratio of $\Delta y$-to-$\Delta x \sim $1.5). Thus, the \citet{jester05} transformations show a similar $(B-R)_0$ and $(B-V)_0$ slope to that of the observed colors, but exhibit a systematic offset along the observed color-color relationship to redder colors (up and to the right in Figure~\ref{fig:BVvBR}).

Both the LVL and \citet{blanton07} transformations show similar agreement between the transformed and observed color slopes in Figure~\ref{fig:BVvBR}, and show small color offsets in Table~\ref{tab:sdss2vegaDelColor}. However, the LVL transformed colors consistently exhibit the lowest, or similar to the lowest, median color differences in all colors; especially those which involve $U-$band transformations.

Figure~\ref{fig:BVvUR} shows another observed color-color trend where there exists a correlation between $(U-R)_0$ and $(B-V)_0$. Note that there is no L05 panel in Figure~\ref{fig:BVvUR} since there are no L05 $U-$band transformations.

Figure~\ref{fig:BVvUR} shows that the \citet{jester05}, \citet{jordi06}, and \citet{blanton07} transformations show a similar slope compared to the observed trend, but the transformed colors of both \citet{jordi06} and \citet{blanton07} show a noticeable offset. Inspection of Table~\ref{tab:sdss2vegaDelColor}  reveals that the \citet{jester05} transformations have redder median colors in both $(U-R)_0$ and $(B-V)_0$ similar to the observed color-color slope indicating a systematic offset along the observed color-color relationship. Furthermore, Table~\ref{tab:sdss2vegaDelColor} shows that all previous transformations exhibit greater median color differences compared to the LVL transformations.

The LVL transformations show good agreement with the color-color trend in Figure~\ref{fig:BVvUR} and no systematic offset in the median color differences in Table~\ref{tab:sdss2vegaDelColor}. In addition, only the LVL transformations show small median color differences in colors which involve $U-$band transformations. Although, $some$ of the previous transformations show good agreement with $some$ observed color trends, only the LVL transformations show consistent agreement across all colors.

\subsection{Johnson-Cousins to SDSS \label{sec:JohnSDSS}}
In this section we derive the $UBVR_c$-to-$ugri$ galaxy transformations with identical methods to those described in \S\ref{sec:SDSStoJohn}. We verify the accuracy of these transformations and compare them to previous transformations via observed $ugri$ color-color trends. However, we only present these results in tabular format since we observe similar trends to those found in \S\ref{sec:SDSStoJohn}.

\subsubsection{LVL Transformations for Galaxies \label{sec:LVLtrans2}}
We examine all possible combinations of SDSS minus Johnson-Cousins magnitudes versus Johnson-Cousins colors to derive the LVL galaxy transformations. The RMS scatter and $\Delta M_{\rm{med}}$ values are calculated for each transformation. The $\Delta M_{\rm{med}}$ versus RMS scatter plot for these transformations (not shown for brevity) has a similar overall structure to that of Figure~\ref{fig:sdss2vegarms}. There are multiple $ugri-$band transformations with low RMS scatter and $\Delta M_{\rm{med}}$ values ($<$0.1 mag). For these bandpasses ($ugri$), we choose the transformations with the lowest RMS and $\Delta M_{\rm{med}}$ added in quadrature (see \S~\ref{sec:LVLtrans}). There are no $z-$band transformations derived in this study due a lack of corresponding $I_c-$band fluxes. The $UBVR_c$-to-$ugri$ galaxy transformation equations and best-fit parameters are presented in Table~\ref{tab:vega2sdss}


\begin{table}
\begin{center}
  {SDSS Transformations}
\end{center}
\begin{tabular}{cccc}
\hline
\hline
Color & & Color Transformation & RMS  \\
      & & 			 & (mag) \\
\hline
$u-V$ & = & $(~~2.05 \pm 0.12)(B-V) + (~~0.10 \pm 0.06)$ & $0.06$\\
$g-V$ & = & $(~~0.70 \pm 0.06)(B-V) + (-0.17 \pm 0.03)$ & $0.05$\\
$r-B$ & = & $(-1.42 \pm 0.08)(B-V) + (~~0.01 \pm 0.04)$ & $0.06$\\
$i-V$ & = & $(-2.29 \pm 0.23)(V-R) + (~~0.48 \pm 0.09)$ & $0.06$\\
\hline
\end{tabular} \\
\caption{The $UBVR_c$-to-$ugri$ transformation equations and coefficients for galaxies. The last column lists the RMS scatter of the LVL galaxy colors around the least $\chi^2$ fit for each transformation.}
\label{tab:vega2sdss}
\end{table}

\subsubsection{Previous Transformations \label{sec:prevtrans2}}
We also compare the LVL $UBVR_c$-to-$ugri$ galaxy transformations to those of previous studies. However, the study of L05 does not provide Johnson-Cousins-to-SDSS transformations, thus there are no comparisons with the L05 study. Furthermore, no $i-$band transformation comparisons can be made with any previous study since these transformations require $I_c-$band photometry which are not available for the LVL sample of galaxies. 

Visual inspection of the previous transformation relationships with LVL galaxy colors overplotted (not shown for brevity) reveals similar results to those seen in Figure~\ref{fig:prevtrans}. The LVL galaxy colors show large scatter around the previous transformation relationships and/or there exists significant deviations between the LVL best fits and those derived in previous studies.

To quantify the accuracy of previous transformations on galaxy colors we calculate the RMS scatter of the LVL galaxy colors around the previously published best fit line and $\Delta M_{\rm{med}}$ values. The results are not graphically shown for brevity but yield similar results to the bottom panel of Figure~\ref{fig:sdss2vegarms}. All of the $g-$band and only one of the $r-$band transformations derived from previous studies yield low RMS scatter and $\Delta M_{\rm{med}}$ values ($<$0.1 mag). However, all of the $u-$ and most of the $r-$band transformations show relatively high RMS scatter ($0.1 < RMS < 0.2$) and $\Delta M_{\rm{med}}$ values ($0.1 < \Delta M_{\rm{med}} < 0.3$). For every transformation of previous studies there exists an LVL transformation with a lower RMS scatter and $\Delta M_{\rm{med}}$ value in every available bandpass.

\subsubsection{Observed Color-Color Trends}
Due to the availability of only three transformed $ugri$ filters, we examine only one observed color-color trend which is not shown for brevity: $(g-r)_0$ versus $(u-r)_0$. We find similar results to those seen in Figure~\ref{fig:BVvUR} where the \citet{jester05}, \citet{jordi06}, and \citet{blanton07} transformations show similar slopes to the observed color-color trends, but the transformed colors of both \citet{jordi06} and \citet{blanton07} show a noticeable offset. Furthermore, the transformed colors of \citet{jester05} show an offset in both colors which are similar to the observed color-color relationship slope which indicates a systematic color offset. Although we do not show this color-color figure, we have provided the median color differences in Table~\ref{tab:vega2sdssDelColor}. 

The LVL galaxy transformations show good agreement with the observed color-color trends and show no systematic color differences in Table~\ref{tab:vega2sdssDelColor}. In addition, the LVL transformations consistently show smaller color differences when compared to previous transformations. These results suggest that the LVL galaxy transformations more accurately describe the SDSS-Johnson-Cousins color relationships for galaxies.

\begin{table}
\begin{center}
  {Median Color Differences}
\end{center}
\begin{tabular}{clc}
\hline
\hline
Color & Study & $\Delta color$ \\
      &       & (mag) \\
$(u-g)_0$.................... & Jester et al. (2005): & $-0.22$\\
                              & Jordi~ et al. (2006): & $-0.08$\\
                              & Blanton \& Roweis (2007): & $-0.12$\\
                              & LVL:~~~~~~~~~~~~~~~ & $~~0.01$\\
\hline \vspace{-0.2 cm}\\
$(u-r)_0$.................... & Jester et al. (2005): & $-0.34$\\
                              & Jordi~ et al. (2006): & $-0.04$\\
                              & Blanton \& Roweis (2007): & $~~0.07$\\
                              & LVL:~~~~~~~~~~~~~~~ & $-0.01$\\
\hline \vspace{-0.2 cm}\\
$(u-i)_0$.................... & LVL:~~~~~~~~~~~~~~~ & $~~0.09$\\
\hline \vspace{-0.2 cm}\\
$(g-r)_0$.................... & Jester et al. (2005): & $-0.10$\\
                              & Jordi~ et al. (2006): & $~~0.01$\\
                              & Blanton \& Roweis (2007): & $~~0.15$\\
                              & LVL:~~~~~~~~~~~~~~~ & $~~0.01$\\
\hline \vspace{-0.2 cm}\\
$(g-i)_0$.................... & LVL:~~~~~~~~~~~~~~~ & $~~0.03$\\
\hline \vspace{-0.2 cm}\\
$(r-i)_0$.................... & LVL:~~~~~~~~~~~~~~~ & $~~0.04$\\
\hline
\end{tabular} \\
\caption{The median color differences between the $UBVR_c$-to-$ugri$ transformed and observed colors for the LVL galaxy transformations and those derived in previous transformations.}
\label{tab:vega2sdssDelColor}
\end{table}

\subsection{Summary}
We have derived empirical color transformations for galaxies between the SDSS $ugri$ and Johnson-Cousins $UBVR_c$ photometric systems. We utilize the LVL nearby galaxy sample which consists of normal, non-starbursting galaxies. The data are taken from the LVL global optical photometry study of Cook et al. (2014a; submitted) where the fluxes were measured within identical apertures across all optical bandpasses to ensure consistent photometric comparisons.  

The LVL galaxy transformations are derived via an analysis of all possible color transformation combinations. The accuracy of each transformation is quantified via the RMS scatter around the least $\chi^2$ fit and the median absolute difference between the resulting transformed and observed magnitudes ($\Delta M_{\rm{med}}$).

We also compared our results to those of previous SDSS transformations derived from standard calibration stars \citep[L05;][]{jester05,jordi06} and model-based galaxy templates \citep{blanton07}. The RMS scatter of LVL galaxy colors around each of the previously published transformations was calculated in addition to the resulting $\Delta M_{\rm{med}}$ value. The observed galaxy colors show large scatter and/or significant offsets for most of the previous color-color relationships. Although $some$ of the previous transformations yielded reasonable RMS scatter and $\Delta M_{\rm{med}}$ values in $some$ filters, there are other filters (i.e., $U-$ and $u-$band) where no previous transformation showed reasonable RMS scatter and $\Delta M_{\rm{med}}$ values ($<$0.15 mag). In addition, for each filter there exists multiple LVL transformations with lower RMS scatter and $\Delta M_{\rm{med}}$ values compared to the best stellar and model-based galaxy transformations.

A secondary check on the accuracy of both the LVL and previous transformations was performed by propagating the transformed magnitudes into colors and comparing them to observed color-color trends. We found that all previous transformations showed significant color offsets ($>$0.1 mag) in more than one color. In general, the model-based galaxy transformations of \citet{blanton07} showed smaller median color differences compared to the stellar transformations, but the LVL transformations consistently showed either the smallest, or similar to the smallest, median color difference in all colors. These results suggest that neither the stellar nor the model-based galaxy transformations can accurately describe all color transformations for observed galaxies when compared to the empirically derived LVL-based galaxy transformations. 

\bibliographystyle{mn2e}   
\bibliography{all}  

\end{document}